\documentstyle[preprint,aps]{revtex}
\begin{document}
\title{ Numerical Evidence of Luttinger and Fermi liquid behavior\\
in the 2D-Hubbard Model }
\author{Sandro Sorella}
\address {International School for Advanced Studies \\
Via Beirut 2-4 I-34013 Trieste Italy}
\date{preprint: SISSA 121/93/CM/MB, BABBAGE: cond-mat/9308001}
\maketitle
\begin{abstract}
The two dimensional Hubbard model with a single spin-up electron interacting
with a finite
density of spin-down electrons is studied using the quantum Monte Carlo
technique, a new
conjugate gradient method for the evaluation of the Edwards wavefunction
ansatz, and the standard
second order perturbation theory.
We performed simulations up to 242 sites at $U/t=4$ reaching the zero
temperature properties
with no ``fermion sign problem'' and found a surprisingly good  accuracy of the
Edwards wavefunction
ansatz at low density or low doping. The conjugate gradient method was then
applied to system
up to 1922 sites and infinite $U$ for the Edwards state.
Fermi liquid theory seems to remain stable in 2D for all cases studied with the
exception of
 the half filling case where a ``Luttinger like behavior'' survives in the
Hubbard model , yielding
a vanishing quasiparticle weight in the thermodynamic limit.
\end{abstract}


\narrowtext

The anomalous properties found in several  High T-c oxides  have renewed
a  considerable attention  and  an increasing interest for  the simplest models
which may explain non conventional  behavior beyond   the Fermi-liquid theory
of normal metals and the BCS theory of usual superconductors..
The Hubbard model is now the most popular hamiltonian
 in condensed matter
physics and  the search for a satisfactory solution  in two spatial dimension
represents one of the most important challenge in theoretical physics.
In one dimension we now have a complete solution, not only for the energy
 spectrum, but also for asymptotic properties of correlation functions.
The physics of the 1D Hubbard model is well described by the Luttinger liquid
 theory. The Fermi liquid theory is unstable in this case due to divergences
in perturbation theory (PT). Such divergences  are  usually not present in
higher
 dimensionality and  a possible anomalous phase in 2D - as suggested by P. W.
Anderson\cite{anderson} - can be explained only
within a  non perturbative approach.

 Historically before  the Lieb-Wu
exact    solution\cite{lieb}   a much simpler  case was solved ; it is the case
when only
 one particle with spin up interacts with a finite density of spin down
 electrons. Although this problem is very much simplified and   probably  far
from reality,   it already contains the basic features
 of 1D conductors : the quasiparticle weight vanishes in the thermodynamic
 limit, and the spectrum consists of holon  and spinon elementary
 excitations\cite{zotos,frenkel}

In this paper we   attempt to search for a non Fermi liquid phase in the 2D
Hubbard model in this simplified sector. In fact although we do not provide an
exact
analytical  solution,   the  numerical advantages will be clearly evident in
this  single
spin- flip Hilbert-space of  the  model.

We thus consider the Hubbard model on a bipartite  lattice , where the
spin-up  and a  finite density $\rho={N\over L}$ of $N$ spin down  electrons
hop
trough the $L$ lattice sites with different hopping  amplitudes, $t_c$ and
$t_d$
respectively. The hamiltonian  therefore reads:
\begin{equation}
H\,=\,  \sum\limits_{R, \tau_{\mu}}
t_c\, c^{\dag}_{R+\tau_{\mu}} c_{R} + t_d\, d^{\dag}_{R+\tau_{\mu}} d_{R}+ U
\sum\limits_{R}  c^{\dag}_R c_R \, d^{\dag}_R d_R
\label{hamiltonian}
\end{equation}
where $d$ ($c$) are fermion operators for spin down (spin up) electrons,
 $\tau_{\mu}$ are the 2d nearest neighbor vectors and d ($=2$ ) is the spatial
 dimension.

We are interested in  the quasiparticle weight :
\begin{equation}
Z_p \,=\, |< \psi_G| d^{\dag}_p |\psi_F>|^2.
\label{zeta}
\end{equation}
where  $|\psi_G>$ is the ground state
of $H$ in the single spin flip subspace and with finite momentum $p$,  and
 $|\psi_F> \,=\, \prod \limits_{\epsilon_k \le \epsilon_F}
 c^{\dag}_k |0>$ is the  non interacting electron  sea of spin down
 electrons, i.e. the ground state of $H$ without   $d-$electrons. Such
a state is assumed in the following to be a non degenerate , translation
invariant closed
 shell state  where $\epsilon_k \,=\, t_c  \sum \limits_{\tau_{\mu}}
 e^{i k\tau_{\mu}}$,and $\epsilon_F$ is the Fermi energy of the down-spin
electrons.  $|\psi_F>$ satisfies these properties  only for particular number
 of electrons, e.g. $N=1,5,13 \ldots  .$, covering all possible densities
 in the thermodynamic limit.

 As well known, the quasiparticle weight
  $Z_p$ measures the strength of the $\delta-$function  in the spectral
 function and
is finite  in the thermodynamic limit if Landau-Fermi liquid theory
is valid.  For example  in  one dimension, where Fermi liquid theory breaks
 down,  $Z$ vanishes
as a power law in the thermodynamic limit
,as predicted by the l Luttinger liquid  theory $Z \propto N^{-\theta}$.
The Hubbard model in the single spin-flip sector is consistent with this
general solution\cite{zotos} and thus represents
 one of the simplest toy models where Fermi-liquid theory
 can be tested in higher dimensions.

The hamiltonian $H$ can be further simplified by tracing out {\em exactly}
 the $d-$electron component from the ground state wavefunction. In fact, using
translation invariance we can write the ground state $|\psi_G>$ of total
 momentum
 $p$ in the following form:
\begin{equation}
|\psi_G>\,=\, {1 \over \sqrt L} \sum\limits_{R}
 e^{-ip R} T_R d^{\dag}_O |\psi>
\label{state}
\end{equation}
 where $|\psi>$ is a wave function
depending only on the spin- down electrons,  $T_R$ is the translation operator
 by a vector $R$ ($T_R\,\, c_{R^{\prime}}\, T_{-R}  \,=\, c_{R+R^{\prime}}$)
 and
$O$ is the origin of the lattice.
Then the following effective hamiltonian $H^{\prime}$ for $|\psi>$ is obtained:
\begin{equation}
H^{\prime}=t_c\, \sum\limits_{R, \tau_{\mu}}
c^{\dag}_{R+\tau_{\mu}} c_{R} \,+\, t_d\,\sum\limits_{\tau_{\mu}} e^{-i p
\tau_{\mu}}\, T_{\tau_{\mu}} + U \,\,c^{\dag}_O c_O .
\label{effhamiltonian}
\end{equation}
  Note that $H^{\prime}$ is not translation invariant and that the  on site
Coulomb
repulsion $U$ becomes  now a simple one-body contribution, i .e. quadratic in
 the
 fermion fields. Using Eq.(\ref{state}) $Z_p$  in Eq.(\ref{zeta})  is replaced
 by the
 overlap  between
 the $U=0$ ($|\psi_F>$ )and the finite $U$ ground state,
 {\em at fixed}   number   of electrons : $Z_p=|<\psi_{F}|\psi >|^2$ .
{\em Thus  the question of Fermi liquid or non Fermi liquid theory
 in this model
is simply related to the stability of the ground state of $H^{\prime}$
under the local perturbation $ U \,\,c^{\dag}_O c_O  $}.

Another useful quantity which we will consider in the following is the momentum
 distribution of the $d-$ electron:
$n_k\,=\, <\psi_G| d^{\dag}_k d_k |\psi_G>. $
By means of Eq. (\ref{state})  $n_k$ is related to the expectation value
of the $p-k$-momentum projector on the state $|\psi>$:
$ n_k \,=\, <\psi| P_{p-k} |\psi >, $ where the projector on the subspace of
 momentum $Q$  is $P_Q \,=\, {1\over L } \sum \limits_R e^{-i Q R } T_R. $
 By inserting in $<\psi| P_{p-k} |\psi >,$ a complete set $|\psi_j>$
 of translation
 invariant states containing  $|\psi_F>$, it easily follows  that:
\begin{equation}
 n_p \ge Z_p
\label{migdal}
\end{equation}
The previous relation can be viewed as a particular case of the Migdal theorem
relating the jump of the momentum distribution at the Fermi surface to the
 amplitude of the spectral weight. In fact we expect that the
inequality (\ref{migdal}) turns in an exact equality in the thermodynamic
limit.

As it is easy to verify the ground state of the hamiltonian
 (\ref{effhamiltonian}) is a free electron
Slater determinant in several limiting cases. For $N=1$ -corresponding to
the two electron problem for the hamiltonian $H$- there is of course no
 correlation in $H^{\prime}$.
 For $t_d=0$
the hamiltonian becomes the well known Falikov-Kimball model and the
 effective hamiltonian $H^{\prime}$  is free and exactly solvable.
 The ground state in  presence of
the local perturbation $U c^{\dag}_0 c_0$ is orthogonal to the non
interacting state, yielding $Z_p \propto N^{-\theta}$ \cite{catastrophe}.
Finally for $U=0$  as well as for $t_d \to \infty$
the free Fermi gas $\psi_F$ is  the ground state and $Z_p=1$.

The above limiting cases are not surprising since  {\em only}
 the term proportional to $t_d$--the spin-up kinetic term--is a  true many body
term in the effective hamiltonian $H^{\prime}$, all remaining ones being
 one-body contributions.  Moreover this spin up kinetic term is obviously not
 extensive in the size of the system  and  can weakly  affect the correlation
in the ground state.
Based on the previous considerations it is likely that the ground state
 of $H^{\prime}$
 is always very close to a simple Slater determinant and thus an Hartree-Fock
(HF)  wavefunction (i.e. the Slater determinant which minimize the expectation
 value for the energy)  may  have a very good overlap with the exact ground
 state of $H^{\prime}$.

The HF wavefunction of the hamiltonian (1) $|\psi_{EWA}>$ is
nothing but the Edwards wavefunction ansatz (EWA),
which is {\em exact} in 1D for $t_{d}=t$  and very accurate in energy
 in the 2D case \cite{edwards}.
Nevertheless  such an Hartree-Fock wavefunction
  corresponds ,by Eq.(\ref{state}) , to a non trivial  {\em correlated }
 state of the Hubbard  hamiltonian $H$. It is  actually a linear
  combination of
$L$- free electron states, yielding for instance the Bethe -ansatz
wavefunction in 1D\cite{edwards,zotos}.

We have used a well known Quantum Monte Carlo (QMC) technique
 \cite{hirsch}
 to evaluate the quasiparticle weight $Z_p$ and the momentum distribution
 $n_p$  in this simple model for $p=0$ and $t_c=t_d$.
. The ground state $|\psi_G>$ is filtered out by imaginary time propagation
 of a given trial wavefunction $|\psi_T>$,
$|\psi_G> \propto \lim\limits_{t \to \infty} e^{-t H} |\psi_T>$, after the
usual
 Trotter-Suzuki decomposition of the imaginary time in $P$ slices of length
$\Delta \tau = {\displaystyle t \over P}$.$|\psi_T> =|\psi_F>$ in all the
 present simulations.  The unrestricted Hartree-Fock calculations for the
 EWA and the straightforward second order perturbation theory for $Z_p$
 and $n_p$ -coinciding at this order-,
 are compared with the QMC
simulations on finite lattices , $l \sqrt 2 \times l \sqrt 2$, with periodic
 boundary conditions tilted by $45$ degrees and
 odd $l$.  The convergence in imaginary time is systematically reached
 within statistical errors in all cases studied. In fact for the
 Hubbard model in the single spin-flip sector the QMC  does not suffer the so
 called ``fermion sign problem'', since in the worst case ($l=11$) the
 average sign is approximately  $0.6$.

In all the QMC simulations I have always found $n_{p=0}$ and $Z_{p=0}$
equals within
statistical errors, with an error bar for $n_p$ three times smaller than
 the one for $Z_p$.
$n_p$ is an upper bound for the quasiparticle weight
(Eq.\ref{migdal}) but  it is always very close to $Z_p$. For instance when
$t_d=0$  or within the EWA approximation $n_0-Z_0 \le 10^{-4}$ for all
 sizes studied.
 In the following we thus identify the two quantities for the  sake of
 simplicity.

We got very accurate results for the evaluation of the EWA wavefunction using
 a {\em new} conjugate
gradients (CG) technique for electronic structure calculations.
In order to apply  the CG algorithm\cite{stich} to the minimization
of an energy
functional $E(\{\psi_n\}) =
 { \displaystyle  <\psi | \,H^{\prime} \,|\psi> \over
 \displaystyle < \psi|\,\psi  > } $,
the orbitals $\psi_n (k)$ of the Slater determinant
 $|\psi>$ have to be  orthogonalized from time to time for numerical
stability\cite{stich}.
In doing so one spoils the efficiency of the conjugate directions,
and the minimization of the energy becomes slower.
This is why  the standard CG algorithm with orthogonalization
does not improve much the steepest descent scheme, as also
discussed in \cite{edwards}.

In order to solve  the previous difficulty we have used the following
 simple scheme.
Since the functional $E(\{\psi_n\}) $ is invariant for any transformation
of the orbitals
$\psi_n  \to \sum_m A_{n,m} \psi_m$ it is possible to  choose the
 $N \times N$ matrix $A$
such that the orbitals read: $\psi_n (k)= \delta_{k,k_n} + u_n (k)$,
 where $k_n$ are wavevectors
 inside the Fermi surface $\epsilon_{k_n} \le \epsilon_F$ and the  functions
$u_n (k)$ all vanish
for $\epsilon_k \le \epsilon_F$.
The number of independent degrees of freedom is thus $N (L-N)$,
as it should be from general ground.
 In this way  we fix the gauge of the transformations, that leave
$E$ unchanged, and we apply the CG strategy, without orthogonalization
 requirements. The method is of course not restricted to the plane
wave basis for the orbitals\cite{io}. In  the largest sizes simulations
($\simeq 2000$ sites) this novel  CG  minimization is approximately
an order of magnitude more efficient  compared to the usual steepest
 descend method. This factor increases with
the size of the system, opening new possibilities for large scale simulations.
 Further details of this new method will be published elsewhere\cite{io}.

At low density we have studied closed shell systems with the  closest
 density just below and above the value
 $\rho={1\over4}$.  This sequence should converge to  $\rho={1\over4}$
 in the thermodynamic limit and should minimize size effects.
In Fig.~\ref{fig1}  we see that size effects are very important and it
is not possible to obtain
 reasonable conclusions with a small size calculation.
The EWA is exact for $N=1$
 (smallest size in  Fig.~\ref{fig1} )  and is a test of our QMC  scheme.
 We note that the results obtained with the  EWA are practically
indistinguishable  from the QMC data, yielding a strong support
 for the accuracy of
the EWA even for relatively large size (up to 242 sites),
 not accessible by exact diagonalization.

Further evidence for the accuracy of the EWA is given by the explicit
 calculation
 of the energy fluctuations $\Delta E^2 \,=\, <\psi|  H^2 |\psi> -
 <\psi|  H |\psi>^2$
 on the EWA $|\psi>=|\psi_{EWA}>$.  $\Delta E^2$ vanishes  for an exact
 eigenstate
 and is given by $U^2 \rho( 1-\rho) $ for the Fermi gas wavefunction
 $|\psi_F>$.
 The Edwards wavefunction typically improves this variance
 by three order of magnitude compared to the Fermi gas wavefunction which still
is very good for this small positive  $U$ value.
As shown in  Fig.~\ref{fig1} the behavior  of the
QMC data  up to $l=11$ and the EWA data for larger systems up to $l=31$ are
 very similar to the
PT, which  is  finite for $d=2$ and $\rho\ne {1\over 2}$.
. Furthermore EWA and QMC data are always well above the PT results
, strongly suggesting that $Z_p$ should be finite in the exact calculation
as well
as for larger $U$ ($U=\infty$ is shown in Fig.~\ref{fig1} ) .
Similar scenario appears also evident for negative
$U$ (Fig.~\ref{fig2}) , corresponding by the particle -hole transformation
\cite{lieb} to positive
 $U$ at density $1-\rho={3\over 4}$ , i.e. the low doping region.
 However in this case the PT
results for $Z_p$ are larger than the QMC and EWA data and the conclusion of
 a finite $Z_p$,
 although quite likely, is less clear. Note also that for infinite $U$,
$Z_p$ seems to drop at the largest sizes.

 At half filling $L/2=N=l^2$ the $U=0$ ground state is a non-degenerate
 closed shell. The Fermi surface is a square and is commensurate with the
finite mesh in the Brillouin zone, leading to size effects very smooth and
well controlled.  In Fig.\ref{fig3}  we show Quantum Monte
 Carlo results as a function of $\Delta \tau ^2$. The
$\Delta \tau ^2\to 0$ limit should be considered a formally exact and unbiased
 property of the ground state for long enough $t$. The results obtained with
 the EWA are systematically larger than the QMC data. except for small sizes.

In this case the second order PT is
logarithmically divergent due to the nesting of the Fermi surface:
$\ln Z_{p=0}  =-U^2 0.00335002 \,\ln(L) + O(U^4).$
 As in 1D this kind of divergence suggests
 a power law decaying for $Z_p \propto N^{-\theta}$. In fact as shown
 in Fig.\ref{fig3}(b) the ln-ln plot appears linear for $\ln(L) \to \infty$
both for the PT data ($\theta_{PT} =0.0536002 $) and for the EWA ones
 ($\theta_{EWA} =0.028$). The QMC data
lie in between the two straight lines giving a strong evidence  that $Z_p$
 eventually vanishes as a power law wit $\theta\sim 0.04.$
This is maybe one of the
 first example of Luttinger-like behavior in a 2D system  and represents the
 central result of this paper. It is also interesting to note that the
 Luttinger liquid exponent for $U=\infty$ is very close to $\theta={1\over 8}$
  for the EWA data, exponent well known in 1D where
it can be determined exactly\cite{parola,schulz}.

In conclusion we have presented an accurate and well controlled size scaling
 of the quasiparticle weight in a numerically tractable sector of the 2D
Hubbard model. We have used QMC  and a {\em new} conjugate gradient
 algorithm for the evaluation of the EWA. This ansatz turns out to be an almost
 exact approximation in most cases for $U=|4t|.$

The quasiparticle weight $Z_p$ looks always finite with the noticeable
exception
of $\rho= {1\over 2}$, where the nesting of the Fermi surface leads to a power
 law decaying $Z_p$, as well as to a logarithmically divergent PT.
At low doping and large $U$  a non perturbative break-down
of Fermi liquid theory is not inconsistent with our data
, and further study is needed to clarify this issue.

I acknowledge useful discussions with A. Parola, M. Fabrizio and E. Tosatti.
 Use of the Cray-YMP  was
 supported by CNR, project ``Sistemi Informatici e Calcolo Parallelo''.

\begin{figure}
\caption{ (a) QMC data (empty dots) vs. EWA ones (full squares) for $U=4t_c$
and $t_c=t_d$. The continuous  line (upper $\rho < 1/4$) connects EWA data for
$N=1,9,21,37$, the lower
curve (dashed $\rho > 1/4$ ) for $N=5,13,25,45,61$.
The QMC data are converged in
imaginary time for $t=l+4$ with $l=3,5,7,9,11$ from left to right.  The
 $\Delta \tau$ correction
was estimated for $l=3,5,7$ with several  points (see Fig.~\protect\ref{fig3}).
(b)  EWA data for $U=4t_c$ (continuous line) and $U=\infty$ ( long dashed line)
  up to $l=31$
compared with second order PT results (dashed line). The arrow indicates
the infinite size PT result. The lines are guides to the eye.}
\label{fig1}
\end{figure}
\begin{figure}
\caption{ same as Fig.1 for negative $U$. The QMC data refers to $l=3,5,7,9$.
Similar considerations apply for the $\Delta \tau$ corrections
and the imaginary time error was negligible  for $t=l+6$.}
\label{fig2}
\end{figure}
\begin{figure}
\caption {
(a) logarithmic plot of the
quasiparticle weight. The empty dots are QMC data  after extrapolation
 to $\Delta \tau \to 0$.
The continuous line and the long dashed line connect EWA data for $U=4t_c$
( full dots)  and  $U=\infty$ (full squares) respectively.
The number of electrons was fixed to $N=l^2$ for $l=3,5,\ldots 23$.
 The PT data (full triangles)  were calculated
up to $l=53$ (not shown in the picture) and the dashed line is the exact
 slope in the thermodynamic limit.
 (b) QMC data as a function of $\Delta \tau^2$ for $U=4t_c$ and
the  imaginary time
$t=l+6$, large enough to have converged results. The continuous lines
are least square fit of the
data and the dashed line ($l=9$) has a slope estimated from the smaller sizes.
 The arrows indicate
the EWA values for $l=3,5,7,9$ from top to bottom figure respectively.}
\label{fig3}
\end{figure}


\begin{references}
\bibitem{anderson} P. W. Anderson, \prl {\bf 64}, 1839 (1990);
{\bf 65}, 2306.
\bibitem{lieb} E. H. Lieb and F. Y. Wu, \prl {\bf 20} 1445 (1968).
\bibitem{zotos} H. Castella and X. Zotos \prb {\bf 47} 16186 (1993).
\bibitem{frenkel} D. M. Frenkel \prb {\bf 46}, 15008 (1992).
\bibitem{catastrophe} P. W.  Anderson, {\it  Phys. Rev.}
{\bf 164} 352 (1967)
\bibitem{edwards}   D.M. Edwards , {\it Progr. Theor. Phys.},
sup. {\bf 101} , 453  (1990); W. von der Linden and D.M. Edwards
 {\it J. Phys. Cond. Matt.} {\bf 3} 4917 (1991).
\bibitem{hirsch} J.E. Hirsch, {\it Phys. Rev. B} {\bf  31}, 4403 (1985);
 S. Sorella {\sl et al.} {\it Europhys. Lett.} {\bf 8}, 663 (1989);
 S.R. White {\sl et al.} {\it Phys. Rev.} {\bf B 40}, 506 (1989).
\bibitem{stich} I. \u{S}tich {\sl et al.} \prb {\bf 39}, 4997 (1989).
\bibitem{io} S. Sorella, in preparation.
\bibitem{parola} A. Parola and S. Sorella  \prl {\bf 64}, 1831 (1990).
\bibitem{schulz} H. Schulz \prl  {\bf 64}, 2831 (1990).

\end{references}
\end{document}